\documentclass[pra,twocolumn,showpacs,floatfix]{revtex4}

\usepackage{amsmath}
\usepackage{times}
\usepackage{graphicx}
\usepackage[dvips]{color}

\begin{document}

\title{Photoassociation spectroscopy of cold alkaline earth atoms near the intercombination line}
\author{R. Ciury\l{}o$^{\dagger\S}$, E. Tiesinga$^{\dagger}$, S. Kotochigova$^{\dagger}$, P. S. Julienne$^{\dagger}$}
\affiliation{$\dagger$ Atomic Physics Division, National Institute of Standards and Technology,
100 Bureau Drive, Stop 8423, Gaithersburg, Maryland 20899-8423, USA.\\
$\S$ Instytut Fizyki, Uniwersytet Miko\l{}aja Kopernika,
Grudzi\c{a}dzka 5/7, 87--100 Toru\'n, Poland.}
\date{\today}

\begin{abstract}
The properties of photoassociation (PA) spectra near the intercombination
line (the weak transition between $^{1}S_{0}$ and $^{3}P_{1}$ states) of
group II atoms are theoretically investigated. As an example we have
carried out a calculation for Calcium atoms colliding at ultra
low temperatures of 1 mK, 1 $\mu$K, and 1 nK.  Unlike in most current
photoassociation spectroscopy the Doppler effect can significantly affect
the shape of the investigated lines.
Spectra are obtained using Ca--Ca and Ca--Ca$^*$ short-range {\it ab initio}
potentials and long-range van der Waals and resonance dipole potentials.
The similar van der Waals coefficients of ground $^{1}S_{0}$+$^{1}S_{0}$
and excited $^{1}S_{0}$+$^{3}P_{1}$ states cause the PA to differ greatly
from those of strong, allowed transitions with resonant dipole interactions.
The density of spectral lines is lower, the Condon points are at relatively short range,
and the reflection approximation for the Franck-Condon factors is not applicable,
and the spontaneous decay to bound ground-state molecules is efficient.
Finally, the possibility of efficient production of cold molecules is discussed.
\end{abstract}


\maketitle

\section{Introduction}

There is a growing interest in the properties of cold alkaline-earth
atoms. One of the main reasons for this interest is a possible
construction of optical clocks, whose precision might exceed
that of the current atomic standard of time\cite{udem02,ma04}.
In particular, optical clocks based on an intercombination transition
of alkaline-earth atoms are seen as good candidates for the next time
standard\cite{wilpers02,curtis03,ido03}.  Increased accuracy of time
standards is, for example, desired for the search for a time-dependent
variation of fundamental constants in atomic experiments, which thereby
would verify claims based on astrophysical data\cite{webb}.

The recent observation of Bose-Einstein condensation (BEC) in an Ytterbium
gas\cite{takasu03} raises hopes for alkaline earth atoms, which have
similar electronic structure. Such achievement
would allow a study of cold gasses over a wide range of temperatures.
Milli- and microkelvin temperatures are reached by Doppler cooling
on the $^{1}S_{0}$--$^{1}P_{1}$ resonance and the $^{1}S_{0}$--$^{3}P_{1}$
intercombination line, respectively.  Nanokelvin temperatures are typical
for Bose condensates and are reached by evaporative cooling.

Another important reason for interest in alkaline-earth atoms is the
absence of a nuclear spin in some isotopes.  This offers an unique
opportunity towards a fundamental study of Doppler cooling. Moreover,
isotopes with zero and nonzero nuclear spin allow a comparison
of Doppler and sub-Doppler cooling\cite{maruyama03,xu03}.
The description of atom-atom interactions is much simpler for a nuclear
spin-less system.  In fact, the basic theory of photoassociation in
strong laser fields\cite{bohn99,simoni02} might be easier to confirm in
alkaline-earth gases than in alkali-metal gases, where to date most
of the research has been done.

Scattering of atoms in ground and excited states lead to a collisional
frequency shift that contributes to the error budget of an optical
clock.  A Bose condensate crucially depends on atom-atom collisions.
Photoassociative (PA) spectroscopy \cite{thorsheim87,lett93,julienne96,weiner99,burnett02}
is one of the most powerful tools to characterize these scattering processes.
It was developed after the success of laser cooling of neutral atoms in
the 1980's \cite{phillips98}.

We will focus our investigation on PA spectroscopy of ultracold
alkaline-earth atoms.  In the presence of laser light, two colliding
ground-state atoms, labelled by the scattering state ``$g$'', absorb a photon
forming an excited molecular bound state ``$e$''\cite{burnett02}. This
process is called photoassociation.  The excited state decays to product
states ``$p$'' leading to detectable loss of atoms from an atomic trap.
The variation of the atom loss as a function of laser frequency gives
the photoassociation spectrum.  The shape of photoassociation lines not
only depends on the properties of the colliding atoms but also on
the temperature and other conditions in a trap\cite{jones99}.

Phoatoassociation spectra close to the resonance of the
$^{1}S_{0}$--$^{1}P_{1}$ transition in alkaline-earth
atoms were theoretically analyzed by Machholm {\it et al.}
\cite{machholm01,machholm02} and others \cite{montalvao01,ribeiro04}.
Recently, Degenhardt {\it et al}. \cite{degenhardt} measured the
photoassociation spectra of cold calcium atoms near this transition
at mK temperatures. Takahashi {\it et al}. \cite{takahashi04}
used the photoassociation spectroscopy to determined the scattering
length of $^{174}$Yb.

This paper analyzes properties of photoassociation spectra near
the intercombination line, i.e. laser frequencies close to the
$^{1}S_{0}$--$^{3}P_{1}$ transition. A dipole transition between pure
singlet and triplet states is forbidden. However, alkaline-earth atom
states labelled $^{3}P_{1}$ are not pure triplet states and have
a small singlet component. This component mostly comes from mixing
with the nearby $^{1}P_{1}$ state and give rise to a weak dipole transition
between the $^{1}S_{0}$ and $^{3}P_{1}$ state.  Consequently, the
$^{1}S_{0}$--$^{3}P_{1}$ atomic line has a small natural width.
As an example we have carried out calculations for calcium.

We describe the shape of photoassociation lines with very-small natural
width and weak laser radiation. The Doppler effect as well as the
photon recoil must be taken into account.  This is in sharp contrast
with the usual treatment of PA\cite{jones99} in which these two
effects are neglected.  Secondly, we discuss
possible patterns of vibrational levels in photoassociation spectra
near the intercombination line. Close-coupled ro-vibrational bound
states are obtained using an interaction Hamiltonian, which is based
on our electronic-structure potentials\cite{kotochigova} and recently
calculated dispersion coefficients\cite{derevianko04}.  The Hamiltonian
also includes coupling between $^{1}P_{1}$ and metastable $^{3}P_{0,1,2}$
states.  The interatomic potential between two ground state calcium
atoms is relatively well known\cite{allard03}.  It is shown that when
the interaction in the ground and excited state is similar the reflection
approximation \cite{bohn99,julienne96,boisseau00} cannot be applied to
calculate the intensities of photoassociations lines. This is unlike
photoassociation spectra near strongly allowed transitions, where the
interaction in the ground and excited states differs significantly and
the reflection approximation is well satisfied.  Finally, we show that
efficient production of cold molecules in the ground electronic state
using photoassociation should be feasible.

\section{Shape of the photoassociation line}

The photoassociation process occurs in a thermal cloud of cold atoms at
temperature $T$ interacting with weak laser radiation.  After absorption
of a photon of frequency $\omega$, two atoms form an excited molecular
bound state $|e\rangle$ with energy $E_{e}$.  The photoassociation process
according to standard descriptions is most efficient when the
photon energy and the kinetic energy of the relative motion of colliding
atoms, $\varepsilon_{r}$, match the energy of the excited bound state,
that is $\hbar\omega+\varepsilon_{r}=E_{e}$.

\begin{figure}
\includegraphics[angle=0,width=\columnwidth,clip]{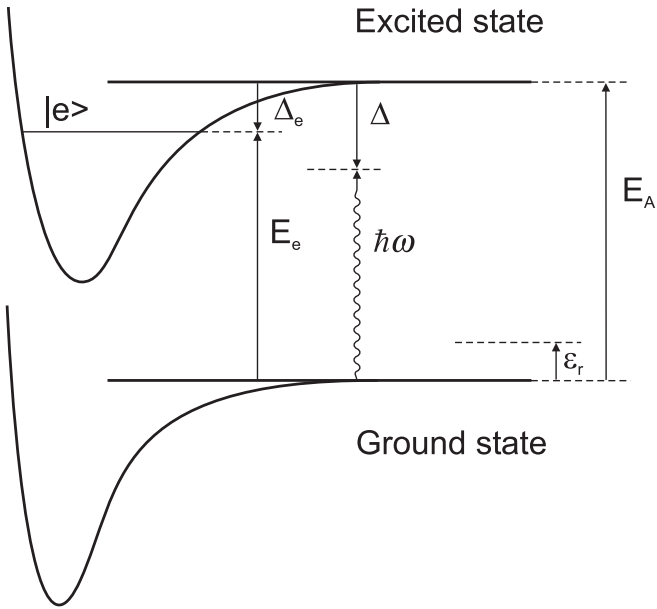}
\caption{Energy diagram of the photoassociation process.
The zero energy corresponds to separated atoms in the electronic ground state
with zero kinetic energy;
$E_{A}$ is the energy of an isolated atom in the excited state;
$E_{e}$ is the energy of the excited molecular bound state $|e\rangle$;
$\hbar\omega$ is the photon energy;
$\varepsilon_{r}$ is the kinetic energy of the relative motion of colliding atoms;
$\Delta_{e}$ is the binding energy of the bound state $|e\rangle$;
$\Delta$ is the detuning of the photon energy from the isolated atom excitation energy $E_{A}$.}
\label{fig1}
\end{figure}

A schematic of energies in the PA process is shown in Fig.~\ref{fig1}.
The binding energy $\Delta_{e}$ of bound state $|e\rangle$ is given
by $\Delta_{e}=E_{e}-E_{A}$, where $E_{A}$ is the energy of an
isolated atom in the excited state and the zero energy corresponds
to separated atoms in the electronic ground state with zero kinetic
energy.  Furthermore, the detuning $\Delta$ of the photon is defined
by $\Delta=\hbar\omega-E_{A}$, where $\hbar\omega$ is the photon
energy. The resonance condition for the PA process in terms of detunings
reads $\Delta+\varepsilon_{r}-\Delta_{e}=0$.

\begin{figure}
\includegraphics[angle=0,width=\columnwidth,clip]{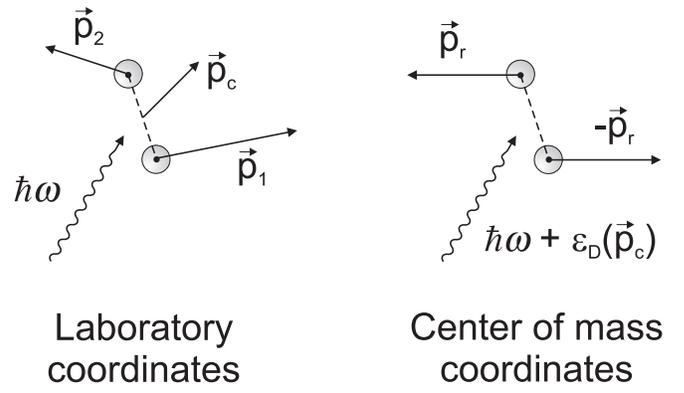}
\caption{Schematic of the collision before the photoassociation process
in the laboratory frame and in the center of mass coordinates.
$\vec{p}_{1}$ and $\vec{p}_{2}$ are momenta of the colliding atoms;
$\vec{p}_{r}$ is the relative momentum of the colliding atoms;
$\vec{p}_{c}$ is the center-of-mass momentum of the two atoms in the laboratory frame;
$\hbar\omega$ is the photon energy in the laboratory coordinates;
$\hbar\omega+\varepsilon_{D}(\vec{p_{c}})$ is the Doppler shifted photon energy
in the moving frame of center of mass coordinates.}
\label{fig2}
\end{figure}

The description of photoassociation for very narrow lines requires us to
include two new effects.  These effects are the Doppler shift and the
photon recoil.  Figure~\ref{fig2} shows a schematic of the collision before the PA
process.  In the laboratory frame the two colliding atoms, each with mass
$m$, have momentum $\vec{p}_{1}$ and $\vec{p}_{2}$, respectively.  In the
coordinate frame, which moves along with the center of mass, the relative
momentum of the colliding atoms is $\vec{p}_{r}$ and kinetic energy of
relative motion $\varepsilon_{r}(\vec{p}_{r})=p_{r}^{2}/2\mu$, where
$\mu=m/2$ is the reduced mass of the colliding atoms.  The photon energy
$\hbar\omega+\varepsilon_{D}(\vec{p_{c}})$ in the moving frame is shifted
with respect to its energy $\hbar\omega$ in the laboratory frame.  To good
approximation the Doppler shift $\varepsilon_{D}(\vec{p_{c}})=-\hbar
\vec{k}_{\rm las}\cdot\vec{p}_{c}/M$ is proportional to $p_{c}$,
where $\vec{p}_{c}$ is the center-of-mass momentum of the two atoms
in the laboratory frame.  The total mass of the system is $M=2m$,
$\vec{k}_{\rm las}$ is the wave vector of the laser radiation with
absolute value $k_{\rm las}=\omega/c$, and $c$ is the speed of light.
After photoassociation the excited molecule gains the momentum
$\hbar\vec{k}_{\rm las}$ of the absorbed photon and, therefore, has
a translational kinetic energy of $E_{\rm rec,mol}=\hbar^{2}k_{\rm
las}^{2}/(2M)$ in the moving frame defined before the absorbtion of
the photon.  Consequently, photoassociation is most efficient when
$\hbar\omega+\varepsilon_{D}(\vec{p_{c}})+\varepsilon_{r}(\vec{p_{r}})
=E_{e}+E_{\rm rec,mol}$ or in terms of detunings
$\Delta+\varepsilon_{D}(\vec{p_{c}})+\varepsilon_{r}(\vec{p_{r}})
-\Delta_{e}-E_{\rm rec,mol}=0$.

The excited molecular states created in the PA process either decay
back to the ground state or can be further excited to ionizing states.
In the former case the product states escape from the trap and give rise
to trap loss.  Ions are detected in the latter case.  In this paper we
will only model trap loss.  The loss mechanisms are characterized by a rate
coefficient $K(\Delta,T)$, which describes the efficiency of the process
for a given laser detuning, intensity $I$, and atom temperature. For
clarity we omit $I$ as argument in the  rate coefficient. The rate
coefficient $K(\Delta,T)$ is linear in the weak laser field intensity $I$.
For higher intensities the center of the PA line shifts linearly with $I$
\cite{bohn99,abraham95,tiesinga96} and the width of the line increases.
The description of the PA line shape presented in this work, for simplicity,
omits this shift as it can be neglected in the weak laser field regime.

The photoassociation trap-loss rate coefficient involves a thermal
average of the rate coefficient for a given pair of momenta $\vec{p}_{1}$
and $\vec{p}_{2}$ in the laboratory frame.  The momentum distribution of
both atoms is Maxwellian with temperature $T$.  In practice we average
over the momenta $\vec{p}_{c}$ and $\vec{p}_{r}$, denoted by
$\left< ... \right>$. These momenta have a Maxwell-Boltzmann distribution
$f_{M}(\vec{p}_{c})=(\sqrt{\pi}p_{M})^{-3}\exp(-p_{c}^{2}/p_{M}^{2})$
and
$f_{\mu}(\vec{p}_{r})=(\sqrt{\pi}p_{\mu})^{-3}\exp(-p_{r}^{2}/p_{\mu}^{2})$
with the most probable momentum $p_{M}=\sqrt{2k_{B}TM}$ and
$p_{\mu}=\sqrt{2k_{B}T\mu}$, respectively, where $k_{B}$ is the Boltzmann
constant. Then, we have
\begin{eqnarray}
\lefteqn{K(\Delta,T)=\left<{\cal K}(\Delta,\vec{p}_{c},\vec{p}_{r})\right>=}
\nonumber\\
&&
\int d^{3}\vec{p}_{c}\, f_{M}(\vec{p}_{c})
\int d^{3}\vec{p}_{r}\, f_{\mu}(\vec{p}_{r})
\; {\cal K}(\Delta,\vec{p}_{c},\vec{p}_{r})
\,,
\label{e1}
\end{eqnarray}
where ${\cal K}(\Delta,\vec{p}_{c},\vec{p}_{r})$  describes the trap loss
from a collision with momenta $\vec{p}_{c}$ and $\vec{p}_{r}$.

The trap-loss coefficient for $\vec{p}_{c}$ and $\vec{p}_{r}$ is equal to
\begin{equation}
\label{e2}
{\cal K}(\Delta,\vec{p}_{c},\vec{p}_{r})=\sum_{e,g}
v_{r}\frac{\pi}{k_{r}^{2}}
|S_{pg}(\Delta,\vec{p_{c}},\vec{p_{r}};e)|^{2}   \,,
\end{equation}
where $k_{r}$ is the relative wavenumber defined by
$\varepsilon_{r}=\hbar^{2}k_{r}^{2}/(2\mu)$ and $v_{r}=\hbar
k_{r}/\mu$ is the relative speed of the colliding atoms.  The quantity
$|S_{pg}(\Delta,\vec{p_{c}},\vec{p_{r}};e)|^{2}$ is the transition
probability from an initial ground state, $g$, to all product states, $p$,
\footnote{It implies a sum over all product states.}
through an intermediate excited bound state $e$. The indices $g$ and $e$,
summed over in Eq.~(\ref{e2}), represent quantum numbers that describe
the initial and intermediate state, respectively. Each initial state $g$
is labeled by the total angular momentum quantum number $J_{g}$, its
projection quantum number $M_{g}$, and the total parity $p_{g}$,
while the intermediate rovibrational levels $e$ are labeled
by vibrational quantum number $v$, total angular
momentum $J_{e}$, projection $M_{e}$ and parity $p_{e}$.

The transition probability from an initial state to the product states
can be described by a generalized resonance formula
\cite{bohn99,julienne96,napolitano94}
\begin{eqnarray}
\lefteqn{|S_{pg}(\Delta,\vec{p_{c}},\vec{p_{r}};e)|^{2}=}
\nonumber\\
&&
\frac{\Gamma_{pe}
\Gamma_{eg}(\varepsilon_{r})}
{(\Delta+\varepsilon_{D}(\vec{p_{c}})+\varepsilon_{r}(\vec{p_{r}})
-\Delta_{e}-E_{\rm rec,mol})^2+(\Gamma_{e}/2)^2} \,,
\nonumber\\
\label{e3}
\end{eqnarray}
where the total width of the excited bound state
$e$, $\Gamma_{e}=\Gamma_{e,\rm nat}+\Gamma_{e,\rm dis}+
\sum_{g}\Gamma_{eg}(\varepsilon_{r})$, is the sum of its natural radiative
width $\Gamma_{e,\rm nat}$ , the contribution $\Gamma_{e,\rm dis}$ from
predissociation, and the stimulated widths $\Gamma_{eg}(\varepsilon_{r})$
caused by the laser coupling between the excited and ground states.
The width $\Gamma_{pe}$ describes the decay into product states
and is assumed to be equal to the sum of the natural radiative and
predissociation width of the bound state.

The stimulated width is proportional to the light intensity
and is calculated from Fermi's golden rule \cite{bohn99,napolitano94}
\begin{equation}
\label{e4}
\Gamma_{eg}(\varepsilon_{r})=2\pi
\left|
\langle\Psi_{e}(v,J_{e}M_{e}p_{e})|V_{\rm las}|\Psi_{g}^{+}(\varepsilon_{r},J_{g}M_{g}p_{g})\rangle
\right|^{2} \,,
\end{equation}
where $|\Psi_{e}(v,J_{e}M_{e}p_{e})\rangle$ is the unit-normalized excited
bound state and $|\Psi_{g}^{+}(\varepsilon_{r},J_{g}M_{g}p_{g})\rangle$ is
the energy normalized scattering ground state.  The operator $V_{\rm
las}$ describes the coupling between the ground and excited state by
laser light.  Details of the close-coupled equations \cite{mies80}
that are solved to calculate the bound and scattering states
are described in appendix \ref{appa}. To describe the atom-photon
interaction during a collision we adopt the treatment developed by
Napolitano {\it et al}. \cite{napolitano97}.
The matrix elements of $V_{\rm las}$ are given in appendix \ref{appb}.

To highlight properties of the photoassociation lines we reduce
the thermal average in Eq.~(\ref{e1}) to the 2D integral
\begin{eqnarray}
\lefteqn{
\left<
\frac{\hbar\pi}{\mu k_{r}}
|S_{pg}(\Delta,\vec{p_{c}},\vec{p_{r}};e)|^{2}
\right>=}
\nonumber\\
&&
\frac{k_{B}T}{hQ_{T}}\;
\frac{2}{\sqrt{\pi}}\int\limits_{-\infty}^{+\infty}dy e^{-y^2}\;
\int\limits_{0}^{\infty}dx\, xe^{-x^2}\;
{\cal L}(\Delta,y,x^{2}) \,,
\label{e5}
\end{eqnarray}
where $y=-\vec{k}_{\rm las}\cdot\vec{p}_{c}/(k_{\rm las}p_{M})$
and $x=p_{r}/p_{\mu}$ are dimensionless variables similar to those
used in the description of pressure and Doppler broadened spectral
lines \cite{harris84,bielski00}, $Q_{T}=(2\pi\mu k_{B}T/h^{2})^{3/2}$, and
\begin{eqnarray}
\lefteqn{{\cal L}(\Delta,y,x^{2})=}
\nonumber\\
&&
\frac{\Gamma_{pe}\Gamma_{eg}(x^{2}\Delta_{T})}
{(\Delta+y\Delta_{D}+x^{2}\Delta_{T}-\Delta_{e}-E_{\rm rec,mol})^2+(\Gamma_{e}/2)^2}
\,.
\nonumber\\
\label{e6}
\end{eqnarray}
The quantities $\Delta_{T}=k_{B}T$ and
$\Delta_{D}=\hbar k_{\rm las}\sqrt{2k_{B}T/M}$ are the thermal
and Doppler width, respectively.

Three limiting cases of the lineshape, Eq.~(\ref{e5}), are of interest.
The shape of the line is Lorentzian when $\Gamma_{e}$ is much bigger
than $\Delta_{D}$, $\Delta_{T}$, and $\Gamma_{eg}(\varepsilon_{r})$.
The denominator of ${\cal L}$ can then be pulled out of the integrals
leading to a Lorentzian profile with a full-width at half maximum
(FWHM) equal to $\Gamma_{e}$.  Such a lineshape can be expected for strongly
allowed transitions at ultra-low temperatures on the order of nanokelvins,
such as exist in Bose condensates \cite{mckenzie02,prodan03} \footnote{In a condensate an
extra factor of 1/2 appears in the rate coefficient expression \cite{stoof89}.}.

In a second limiting case the shape of the line is a ``cut-off
exponential''.  This profile can be obtained when $\Delta_{T}$ is much
larger than $\Delta_{D}$ and $\Gamma_{e}$ and the energy dependence
of $\Gamma_{eg}(\varepsilon_{r})$ is neglected
(compare discussion in Ref. \cite{jones99} where the energy dependence
$\Gamma_{eg}(\varepsilon_{r})\sim\varepsilon_{r}^{l_{g}+1/2}$ describing
the Wigner threshold law behavior is considered).
The Lorentzian ${\cal L}$ can be replaced by a delta-Dirac function with
argument $\Delta-\Delta_e-E_{\rm rec,mol}+x^2\Delta_T$ and the integrals
can be solved analytically. In fact, the profile is proportional
to $\theta(-\Delta)\exp(\Delta/\Delta_{T})$, where $\theta(z)$ is the
Heaviside step function: $\theta(z)=1$ for $z\ge1$ and $\theta(z)=0$ for $z<1$.
The full width at $1/e$ of the exponential
lineshape equals the thermal width $\Delta_{T}$. This lineshape is most
easily observed at magneto-optical trapping temperatures on the order
of a millikelvin.

Finally, for the unusual situation of extremely weak transitions at
nanokelvin temperatures one could try to achieve conditions in which
$\Delta_{D}$ is much bigger than $\Delta_{T}$ and $\Gamma_{e}$. In such
a case the Lorentzian can again be replaced by a delta function but now
with an argument that only depends on $y$ and $\Delta$. The resulting
lineshape is a Gaussian with half width at $1/e$ of the maximum equal
to $\Delta_{D}$.

In the usual treatment of the PA line shape Doppler broadening is
neglected. To find conditions for which this approximation breaks
down, the relative importance of Doppler and thermal effects must
be determined. It is easy to show that $\Delta_{D}=\Delta_{T}$ at
temperature $T=T_{R}$, where $T_{R}=\hbar^2 k_{\rm las}^{2}/(mk_{B})$
is the atomic recoil temperature.  At temperatures $T>T_{R}$ thermal
broadening dominates, while for $T<T_{R}$ Doppler broadening can determine
the shape of the line. In fact, this requirement is not sufficient. It is
also necessary to assume that the Doppler width is comparable or bigger
than $\Gamma_{e}$.

The influence of the photon recoil on the PA spectra is to very good
approximation described by an uniform shift, $E_{\rm rec,mol}$,
of all PA lines. It should be noted that the photon recoil energy
of the two-atom molecule is two times smaller than the photon recoil
energy of an isolated atom $E_{\rm rec}=\hbar^2 k_{\rm las}^{2}/(2m)$.
This indicates that for molecular bound states close to atomic
levels there should exist a transition region from recoiling as a molecule
to recoiling as an atom. Our theory does not treat this effect and
should only be applied to molecular states with a binding energy
that is much bigger than the photon recoil energy.

\section{Interatomic Hamiltonian}

Properties of photoassociation spectra are governed by the interactions
between colliding atoms. The interaction Hamiltonian is described
in appendix \ref{appa} and is similar to that discussed by Mies {\it et
al.}\cite{mies78} for the electronic structure and spectroscopy of
Hg$_{2}$.  This Hamiltonian includes non-relativistic Born-Oppenheimer
potentials, the spin-orbit splitting of the $^3P_{0,1,2}$ atomic states,
relativistic coupling between $^1P$ and $^3P$ states,
and a term that incorporates the rotation of the two atom system.

The best possible Born-Oppenheimer potentials are used.  {\it Ab initio}
calculations, with the exception of a few simple cases, do not give
sufficiently accurate predictions of absolute positions of molecular
bound states. Therefore theoretical potentials are used as initial
guesses, and modified at short interatomic separation
to reproduce experimental binding energies.
Unfortunately there is no experimental data on binding energies
near the $^3P_1+{}^1S_0$ dissociation limit for the calcium molecule.
As the rovibrational structure is not known experimentally, we can only
map out possible spectra by varying the short-range
part of the potential.

The Born-Oppenheimer potentials have Hund's case (a) symmetry,
$^{2S+1}|\Lambda|_{\sigma}$, where $S$ is the total electron spin, $\Lambda$
is the projection of the total electron orbital
angular momentum along the interatomic axis, and $\sigma=g/u$ describes
the gerade or ungerade symmetry of a state.  Near the $^3P_1+{}^1S_0$
limit, where the atomic spin-orbit interaction is much bigger than the
Born-Oppenheimer potentials, adiabatic potentials are better described by
Hund's case (c) symmetry, $|\Omega|^{\pm}_\sigma$, where $\Omega$ is the
projection of the total electron angular momentum along the interatomic
axis and for $\Omega=0$, the label $\pm$ describes the symmetry under
a reflection of the electronic wavefunction.
The adiabatic Hund's case (c) potentials are obtained by simultaneous
diagonalization of the Born-Oppenheimer potentials and the spin-orbit coupling.

\begin{figure}
\includegraphics[angle=0,width=\columnwidth,clip]{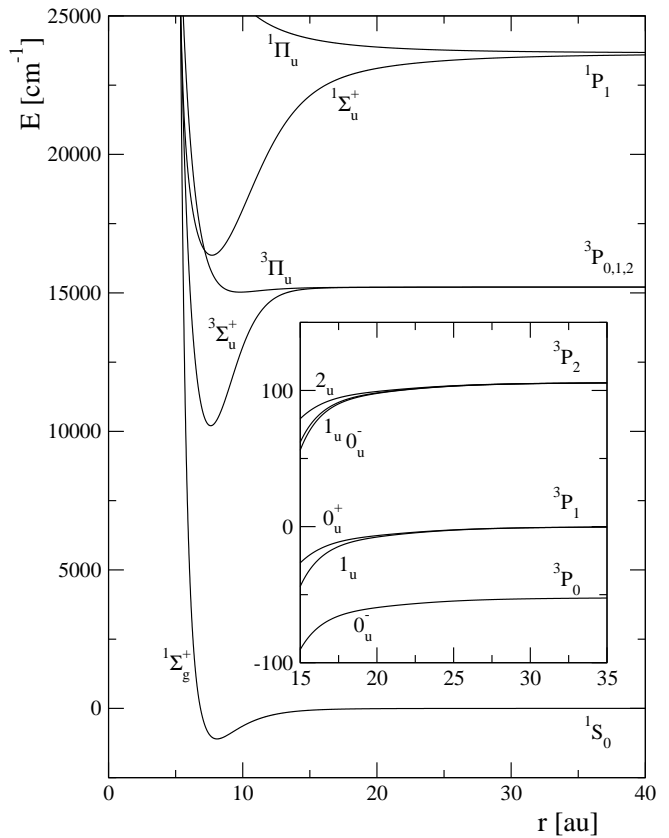}
\caption{Adiabatic potentials of two calcium atoms as a function
of interatomic separation.  The inset shows a blowup of the region near
the $^3P+{}^1S$ limits.  At short range the potentials are labelled by
their Hund's case (a) symmetry and in the inset by a Hund's case (c)
symmetry label. Ungerade excited states, accessible by an optically allowed
transition from the ground X$^1\Sigma_g^+$ potential, are shown.}
\label{fig3}
\end{figure}

Relevant data about the interaction potentials between alkaline-earth
atoms in the ground and excited states have been compiled by Kotochigova
and Julienne \cite{kotochigova}.  The Ca$_{2}$ potentials  \cite{kotochigova}
are shown in Fig.~\ref{fig3}. Czuchaj {\it et al}. \cite{czuchaj03} have
published a similar set of potentials.
It is convenient to use atomic units.  The atomic unit
of length $1\;{\rm au}=1\;a_{0}=0.0529177\;{\rm nm}$ is equal to a
Bohr radius $a_{0}=\hbar^{2}/(m_{e}e^{2})$; the atomic unit of energy
$1\;{\rm au}=E_{h}=4.3597482\times10^{-18}\;{\rm J}$ is equal to the
Hartree energy $E_{h}=\hbar^2/(m_{e}a_{0}^2)$. Here $m_e$ and $e$ are
the electron mass and charge, respectively.

The $X^{1}\Sigma_{g}^{+}$ electronic ground state potential of Ca$_2$
has been determined by  Allard {\it et al}.\cite{allard03}.  Parameters of
the potential are listed in Table 1 of this reference. For large
interatomic separations the potential asymptotically approaches
a van der Waals potential with $C_{6}(X^{1}\Sigma_{g}^{+})=
2081.18\;{\rm au}$.  The atomic unit of $C_6$ is $1\;{\rm
au}=E_{h}a_{0}^{6}=0.957342\times10^{-79}\;{\rm Jm^{6}}$. The scattering
length for this potential is $a_{\rm scat}=389.8\;{\rm au}$.

The exited-state interaction potentials between calcium atoms dissociating
to the $^{1}P_{1}+{}^{1}S_{0}$ and $^{3}P_{0,1,2}+{}^{1}S_{0}$ limits
for short-range interatomic separations have been modelled using the
adiabatic $^{1,3}\Sigma^+_{g,u}$ and $^{1,3}\Pi_{g,u}$ potentials.
These adiabatic potentials are determined on the basis of {\it ab
initio} calculations of Ref.~\cite{kotochigova} and
smoothly connected to their asymptotic functional form.  The form is
$C_{6}/r^{6}$ for triplet states and $C_{3}/r^{3}$ for singlet states,
respectively. Triplet potentials dissociate to the $^{3}P+{}^{1}S$ limits,
while singlet potentials dissociate to the $^{1}P+{}^{1}S$ limit.

In the model describing interaction of two calcium atoms we do not include
potentials correlating to $^{1}D_{2}+{}^{1}S_{0}$ \cite{bussery03}
and $^{3}D_{1,2,3}+{}^{1}S_{0}$ dissociation limits.
These potentials could give rise to molecular bound
states near the $^{3}P_{1}+{}^{1}S_{0}$ limit, but the sparse density of states
makes it unlikely that such level occur in a small energy interval
close to the $^{3}P_{1}+{}^{1}S_{0}$ dissociation limit.

Dispersion coefficients for two ground state atoms are well known
for many atomic species \cite{allard03,prosev02,mitroy03}. For excited
atoms, however, there is little data.  For a calcium atom in the
ground $^{1}S$ state interacting with another calcium atom in
the excited $^{3}P$ state the long-range dispersion coefficients
have been recently calculated by Derevianko and Porsev\cite{derevianko04}.
The attractive Hund's case (c) $|\Omega|^{\pm}_{g/u}$ potentials
correlating to the $^{3}P_{1}+{}^1S_0$ limit have
$C_{6}(0^{+}_{g/u})= 2462\;{\rm au}$
and $C_{6}(1_{g/u})= 2593\;{\rm au}$ \cite{derevianko04}.
For this Paper, however, dispersion coefficients are needed
for Hund's case (a) $^{2S+1}|\Lambda|_{g/u}$
Born-Oppenheimer potentials. Following Ref.~\cite{mies78}
we have $C_{6}(0^{+}_{g/u})=C_{6}(^{3}\Pi_{g/u})$,
$C_{6}(1_{g/u})=[C_{6}(^{3}\Sigma_{g/u})+C_{6}(^{3}\Pi_{g/u})]/2$,
and, therefore, $C_{6}(^{3}\Sigma_{g/u})= 2724\;{\rm au}$,
$C_{6}(^{3}\Pi_{g/u})= 2462\;{\rm au}$.
Derevianko and Porsev\cite{derevianko04} found that the splitting between
the long-range $\Sigma$ and $\Pi$ potentials is small.
This small anisotropy for the quasi-two-electron atom is qualitatively
different from analytical predictions for single-electron excited alkali-metal
atoms interacting with rare-gas atoms\cite{mahan68,mahan69}.
Older data \cite{merawa01} does
not agree with the results of Ref. \cite{derevianko04}.

The lifetime of the atomic $^{1}P_{1}$ state determines
the dipole-dipole interaction coefficients, $C_3$, of the
singlet potentials.  In fact, $C_{3}(^{1}|\Lambda|_{\sigma})=
\zeta(^{1}|\Lambda|_{\sigma}) C_{3}^{(0)}$ where $C_{3}^{(0)}=
(3/4)\Gamma_{A}(^{1}P_{1}) [\lambda_{A}(^{1}P_{1})/(2\pi)]^{3}$,
$\Gamma_{A}(^{1}P_{1})=\hbar/\tau_{A}(^{1}P_{1})$ is the natural width of
the excited $^{1}P_{1}$ state, and $\lambda_{A}$ is the wavelength of the
corresponding radiation.  The coefficient $\zeta(^{1}|\Lambda|_{\sigma})$
is defined by $\zeta(^{1}\Sigma_{g})=+2$, $\zeta(^{1}\Sigma_{u})=-2$,
$\zeta(^{1}\Pi_{g})=-1$, and $\zeta(^{1}\Pi_{u})=+1$.
For calcium $\tau_{A}(^{1}P_{1})= 4.59\;{\rm ns}$ \cite{machholm01,wiese}.

Retardation effects\cite{meath68,power67} do not change our main
conclusion.  The calcium bound states of interest are mostly confined to
interatomic separations that are small compared to the
wavelength $\lambda_{A}(^{1}P_{1})$.  Under such circumstances retardation
effects can be neglected. Moreover, this implies that photoassociation
of two ground state atoms can only excite the ungerade states. Therefore,
Fig.~\ref{fig3} only shows ungerade excited potential curves.

\section{Results}

Photoassociation spectra near the $^{3}P_{1}+{}^{1}S_{0}$ limit
are expected to be weak because the atomic transition dipole to
the intercombination line is nearly forbidden.  As we will show,
such spectra can most easily be measured at ultracold temperatures on the
order of $\mu{\rm K}$ and below.  At these temperatures only $s$-wave
collisions will contribute to the spectrum.  Consequently, contributing
transitions are between the ground {\it gerade} scattering state,
$|\Psi_{g}^{+}(\varepsilon_{r},J_{g}M_{g}p_{g})\rangle$,  with total molecular
angular momentum $J_{g}=l_{g}=0$ and parity $p_{g}=1$, where $l_{g}$ is the rotational angular
momentum between the atoms, and excited {\it ungerade} bound states
$|\Psi_{e}(v,J_{e}M_{e}p_{e})\rangle$ with $J_{e}=1$ and $p_{e}=-1$.
For total molecular angular
momentum $J_{g}=0$ and $J_{e}=1$ there are one and five coupled channels,
respectively (See Appendix \ref{appa}.).  The excited-state channels for
odd $J_{e}$ and negative total parity $p_{e}$ do not
include the $^{3}P_{0}$ atomic state and, therefore, the predissociation
width $\Gamma_{e,\rm dis}$ in Eq.~\ref{e4} is zero
\footnote{Excited bound states with even $J_{e}$ and negative parity
can predissociate to a scattering state dissociating to the
$^{3}P_{0}+{}^{1}S_{0}$ limit.
Such a level can only be excited from $J_{g}=l_{g}\ge 2$ ground
scattering states and the excitation probability is negligible at ultracold
temperatures on the order of $\mu{\rm K}$ and below.}.

\begin{figure}
\includegraphics[angle=0,width=\columnwidth,clip]{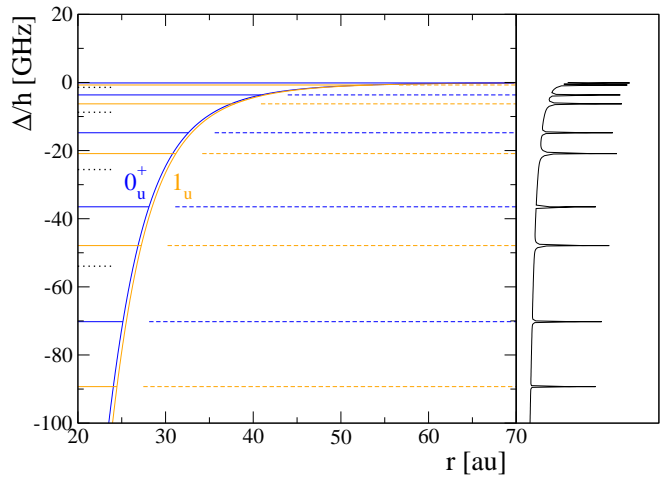}
\caption{An example bound-state structure and corresponding
photoassociation spectrum of calcium near the intercombination line.
The detuning $\Delta$ is defined relative to the $^{3}P_{1}+{}^{1}S_{0}$
limit.  The panel on the left shows the $0_u^+$ (dark line) and $1_u$
(light line) Hund's case (c) adiabatic potentials as a function of
interatomic separation $r$. The bound states of these potentials are indicated by
the thick and thin solid horizontal lines, respectively.
The panel on the right shows the corresponding photoassociation spectrum under
typical conditions. Each peak corresponds to a bound state in the left panel
as indicated by the dashed lines. Finally, the short dotted lines on the
left side of the graph indicate ``bin edges'' of the $0_u^+$ potential.}
\label{fig4}
\end{figure}

Figure~\ref{fig4} shows an example of a bound state structure and the corresponding
spectrum.  The PA spectrum is a reflection of the rovibrational structure
of excited molecules.  In this case lines are assigned to the Hund's case (c)
$0_{u}^{+}$ or $1_{u}$ symmetry.

Details of the Born-Oppenheimer potential at short interatomic separation
are insufficiently known for a quantitative prediction of the bound-state locations.
Therefore, we have modified the short range of the $^{3}\Pi_{u}$
potential to demonstrate how a spectrum can change. The $0_{u}^{+}$ and $1_{u}$
potentials correlate to the $^{3}\Pi_{u}$ and $^{3}\Sigma_{u}^{+}$ potentials
at short range, respectively. Consequently, by changing
the $^{3}\Pi_{u}$ potential we can change the location of the $0_{u}^{+}$ bound states
while leaving the position of $1_{u}$ bound states virtually unchanged.
It is convenient to define ``bins'' of the $0_u^+$ potential as energy
intervals with edges marked by the $0_{u}^{+}$ energy levels calculated
with a $^{3}\Pi_{u}$ potential, such that the last bound state is exactly on
the $^{3}P_{1}+{}^{1}S_{0}$ threshold. As the short-range $^{3}\Pi_{u}$
potential is changed, there is always exactly one $0_{u}^{+}$ bound state
in each bin. The ``bin edges'' for $0_{u}^{+}$ states are shown in Fig.~\ref{fig4}.

\begin{figure}
\includegraphics[angle=0,width=\columnwidth,clip]{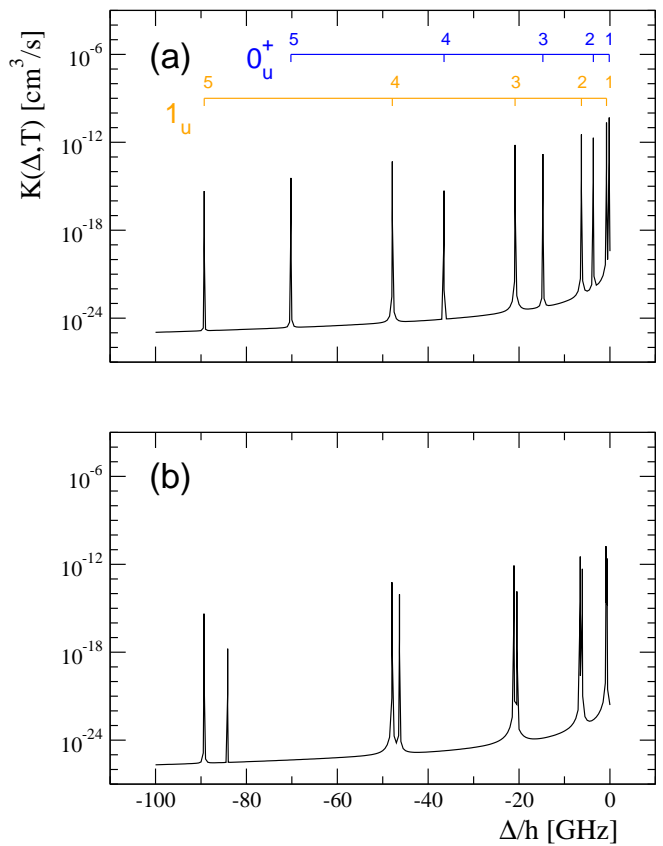}
\caption{Photoassociation spectra of calcium near the intercombination
line for two $^{3}\Pi_{u}$ potentials. Panel (a) is for the potential
used in Fig.~{\protect\ref{fig4}}.  Panel (b) is for a potential where the
$0_u^+$ and $1_u$ bands nearly overlap.  The laser intensity is 1 W/cm$^2$
and the temperature of the gas is $T=1$ $\mu$K.  The $0_{u}^{+}$ and $1_u$
vibrational assignment is shown in Panel A.  The last vibrational level
is labeled by ``1''. }
\label{fig5}
\end{figure}

Figure~\ref{fig5} shows PA spectra for two $^{3}\Pi_{u}$ potentials in
order to illustrate the limiting cases of overlapping and non-overlapping
$0_{u}^{+}$ and $1_{u}$ bands. The PA spectra are calculated using
the lineshape formula derived in the previous section and take into
account Doppler broadening.  In both spectra the last five vibrational
levels of $0_{u}^{+}$ and $1_u$ symmetry are shown.  The PA rate
coefficient increases by nine orders of magnitude near each vibrational
level. Observable trap loss is on the order of the $10^{-12}\;\rm cm^{3}
s^{-1}$.  In principle, ion detection allows measurement of weaker PA
lines than is possible with trap loss. On the frequency scale of the
figure the location of the $1_u$ vibrational lines for the two potentials
is almost the same.

A spectrum in which lines with $0_{u}^{+}$ and $1_u$ symmetry are far
apart from one another is shown in Fig.~\ref{fig5}(a). In this case
the projection $\Omega_{e}$ of the electronic angular momentum $j_{e}$
on the intermolecular axis is a good quantum number and bound states
can be labelled by the Hund's case (c) coupling scheme. This kind of
spectrum has been observed in preliminary experiments on Strontium near
the $^{1}S_{0}$--$^{3}P_{1}$ line by Ido and Katori \cite{ido01}.

The rather small anisotropy of the long-range dispersion interaction
of the $0_{u}^{+}$ and $1_u$ potentials ($C_{6}(0_{u}^{+})\approx
C_{6}(1_{u})$ to within 5\%) can lead to a near coincidence of
$0_{u}^{+}$ and $1_{u}$ levels over a range of $v$. Such a case is
shown in Fig.~\ref{fig5}(b). For closely spaced doublets the projection
$\Omega_{e}$ is not a good quantum number, and bound states should
rather be labelled by the rotational angular momentum $l_{e}$ as in
the Hund's case (e) coupling scheme. It can be clearly seen that in
each doublet there is a strong and weak line corresponding to $l_{e}=0$
and $l_{e}=2$, respectively.

We have also studied the change in coupling scheme from Hund's case (c) to
$e$ for bound states very close to the molecular thresholds.  By changing
the shape of the $^{3}\Pi_{u}$ potential, the last bound state, initially
attributed to $0_{u}^{+}$ symmetry, smoothly approaches threshold.
Simultaneously, the wave function smoothly changes its character as well.
For a binding energy $\Delta_{e}/h=-0.026\;{\rm GHz}$ 90\% of the wave
function has $\Omega_{e}=0$ character, while for $\Delta_{e}/h=-0.003\;{\rm
GHz}$ 90\% of the wave function can be attributed to $l_{e}=0$.
In other words, Hund's case (c) holds for binding energies larger than
0.026 GHz.

In Fig.~\ref{fig5} and in the remainder of this paper the natural
linewidth $\Gamma_{e,\rm nat}$ is approximated by $2\Gamma_{A}(^{3}P_{1})=0.663$ kHz.
The natural linewidths calculated from the theory described in
the appendices shows that the width is less than about four times
$\Gamma_{A}(^{3}P_{1})$ for the detunings shown in Fig.~\ref{fig5}.
We believe that our simple model can not quantitatively describe the
natural widths. The treatment of the coupling between $^1P$ and $^3P$
states is not sufficiently accurate. For simplicity we assume the
same natural linewidth for all lines.

\begin{figure}
\includegraphics[angle=0,width=\columnwidth,clip]{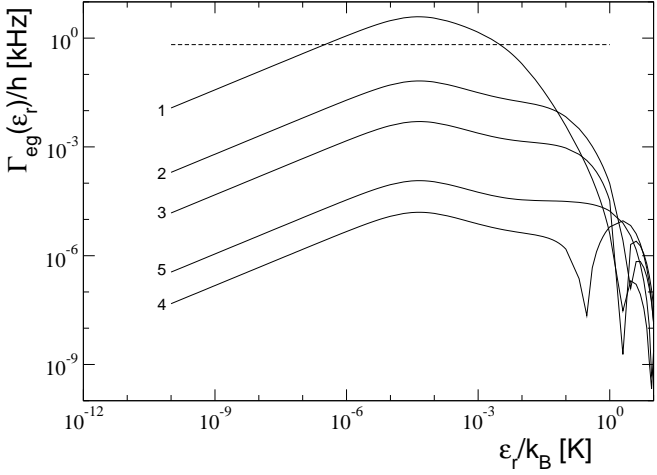}
\caption{The stimulated width as a function of the kinetic energy of a
$^1S+^1S$ collision calculated for the five lines of the $0_u^+$ band shown in
panel (a) of Fig.~{\protect\ref{fig5}}.  The laser intensity is 1 W/cm$^2$.
The natural width $\Gamma_{e,\rm nat}/h=0.663$ kHz is marked by the dashed line.}
\label{fig6}
\end{figure}

Figure~\ref{fig6} displays the collision-energy dependence of the
stimulated width $\Gamma_{eg}(\varepsilon_{r})$ for the $0_{u}^{+}$
lines shown in Fig.~\ref{fig5}(a). The ground-state potential has
a scattering length of 389.8 $a_0$ and the laser intensity is 1
W/cm$^2$. The width rapidly increases for bound states closer to the
threshold as the overlap of the bound and scattering wavefunction grows.
Moreover, for collision energies smaller than $\varepsilon_r/k_B=100$
$\mu$K the width of all lines is proportional to $\sqrt{\varepsilon_r}$,
satisfying the Wigner threshold law.  For a 1 W/cm$^2$ laser intensity
and collision energies less than 1 $\mu$K the stimulated width for four
of the vibrational levels is smaller than the natural width.  Therefore,
the lines are unsaturated for most of lines shown in Fig.~\ref{fig5}.
The exception are those lines closest to resonance.

\begin{figure}
\includegraphics[angle=0,width=\columnwidth,clip]{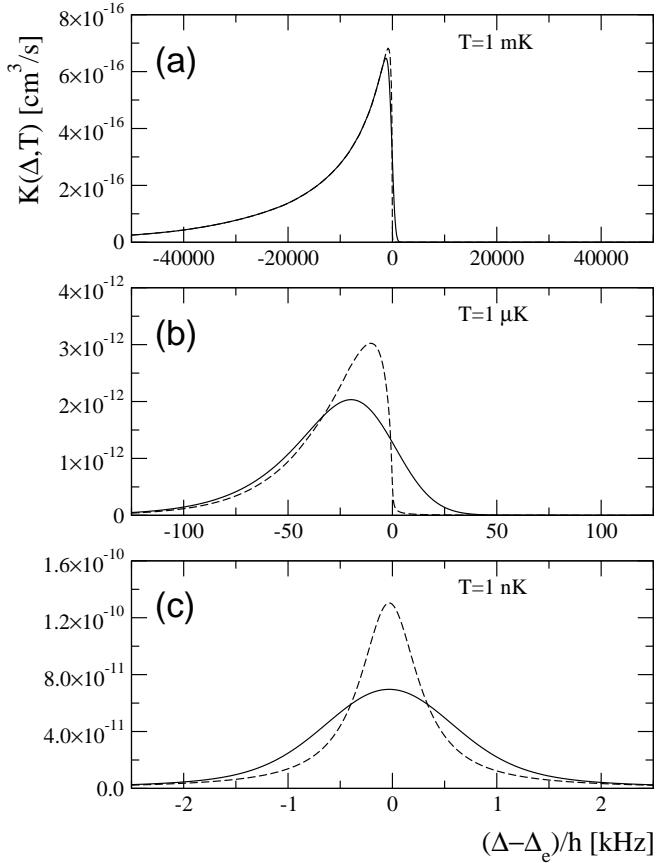}
\caption{The lineshape of the $0_u^+$ line labelled 2 shown in panel (a) of
Fig.~{\protect\ref{fig5}} as a function of laser detuning.  Panels (a),
(b) and (c) show lineshapes at a temperature of 1 mK, 1$\mu$K and 1 nK,
respectively.  Profiles that include and do not include Doppler broadening
are marked by solid and dashed lines, respectively.  The laser intensity
is 1 W/cm$^2$.  The quantity $\Delta_e$ is the binding energy of the
$0_u^+$ rovibrational level relative to the $^1S_0+^1P_1$ dissociation
limit.}
\label{fig7}
\end{figure}

In order to show the temperature dependence of the shape of a line we
have chosen line 2 of the $0_{u}^{+}$ band in Fig.~\ref{fig5}(a).
Figure~\ref{fig7} shows the lineshape for a temperature of 1 mK, 1 $\mu{\rm K}$,
and 1 nK with and without Doppler broadening. In the absence
of Doppler broadening $\Delta_D=0$ in Eq.~\ref{e6}.  For these three
temperatures the thermal width $\Delta_{T}/h=$ 20837 kHz, 20.8 kHz,
and 0.021 kHz and the Doppler width $\Delta_{D}/h=$ 694 kHz, 21.9 kHz,
and 0.694 kHz, respectively.

At a temperature of 1 mK the line in Fig.~\ref{fig7}(a) has the typical
``cut-off exponential'' shape determined by thermal broadening and is
only slightly affected by Doppler broadening. The width of line is on
the order of ten MHz and the peak rate coefficient is $10^{-15}\;\rm
cm^{3}s^{-1}$. Such a low rate coefficient makes trap loss hard to
detect in typical ultra-cold-atom experiments in a magneto-optical trap.
In this case ion detection might be a sensitive alternative.

A 1 $\mu$K atomic-gas temperature is close to the recoil temperature
$T_{R}=1.11\;{\rm \mu K}$.  Under such conditions both thermal and Doppler
broadening in the PA lineshape are comparable.  Figure~\ref{fig7}(b)
demonstrates a significant difference between a Doppler broadened
profile and one without Doppler broadening.  The width of the line
is on the order of hundred kHz and the peak rate coefficient is $10^{-12}\;\rm
cm^{3}s^{-1}$. A trap-loss signal should be observable for such rate
coefficients.

In Fig.~\ref{fig7}(c) a PA lineshape for a thermal gas at 1 nK is shown.
Typically for such low temperatures and sufficiently high densities an
atomic gas could be Bose condensed(BEC) and a Boltzman distribution
of atomic momentum should not be used.  Here, we assume a low enough
density that condensation has not occurred.  The lineshape is an ordinary
Voigt profile, which is determined by Doppler and natural broadening.
The natural width $\Gamma_{e,\rm nat}$ is 0.663 kHz.  The line is
Lorentzian if Doppler broadening is neglected.  The width of the line is
on the order of kHz and the peak rate coefficient is $10^{-10}\;\rm cm^{3}s^{-1}$.
Moreover, the molecular recoil energy $E_{\rm rec,mol}/h=$ 5.775 kHz is
significantly bigger than the width of the line.

The width of the lines in Fig.~\ref{fig7} varies by four orders of
magnitude.  The peak rate coefficient changes by five orders of magnitude.  Clearly,
for temperatures on the order of 1 $\mu{\rm K}$ and below, photoassociation
spectra should be observable.  Moreover, Doppler broadening is an
important factor and affects the shape of the lines significantly.

\begin{figure}
\includegraphics[angle=0,width=\columnwidth,clip]{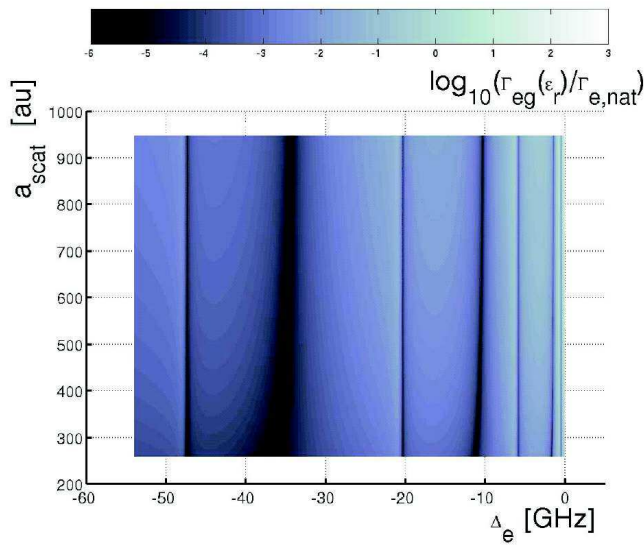}
\caption{The stimulated line width of
the $0_u^+$ band as a function of their line position (binding energy) $\Delta_{e}$
and scattering length $a_{\rm scat}$ of the ground state potential.
The laser intensity is 1 W/cm$^2$ and the kinetic energy
of the collision is $\varepsilon_{r}/k_{B}=1\;\mu{\rm K}$.}
\label{fig8}
\end{figure}

We have also analyzed the stimulated width of the $0_u^+$ lines as a
function of the scattering length in the $^1S_0+^1S_0$ ground state and
the binding energy $\Delta_e$ of the excited $0^+_u$ bound states.  The scattering
length is varied within the range allowed by experiment\cite{allard03}.
The scattering length is varied by slight modifications of the
short-range part of the ground-state potential. The binding energy of
the $0_u^+$ bound states are changed by modifying the $^3\Pi_u$ potential
as discussed above.

Figure~\ref{fig8} shows the stimulated width at a collision energy
$\varepsilon_r/k_B=1$ $\mu$K as a function of $a_{\rm scat}$ and
$\Delta_e$.  The figure shows multiple nearly vertical dark structures,
where the width is nearly zero.  There are two kinds of these structures:
ones accompanied by a parallel bright feature, where $\Gamma_{eg}$ is
large, and those without.  The latter structures are not quite vertical
for smaller $a_{\rm scat}$.  The ``first kind'' of structure occurs when a
$0_u^+$ bound state coincides with a $1_{u}$ bound state.  This mixing
is independent of any ground state scattering property and therefore
the structures are vertical.
The ``second kind'' of dark structures occur when the overlap of the scattering
wavefunction and the excited bound state vanishes. The shape of the
scattering wavefunction near the outer turning point of the excited bound
state does not change much when the scattering length is on the order of
a few hundred $a_{0}$.  Therefore, these structures are nearly vertical.
Near $a_{\rm scat}=300$ $a_0$ a departure from vertical can be observed.

For photoassociation near strongly-allowed atomic transitions, such as
occur in alkali-metal experiments\cite{jones99,weiner99} and near the
$^{1}P_{1}$ line of the alkaline earths\cite{machholm01}, the overlap
vanishes for excited bound states with outer turning points near the nodes
of the ground-state wave function. Their detuning $\Delta_e$ can be
found with the help of the reflection approximation\cite{julienne96,boisseau00},
which says that the Franck-Condon factor is proportional to the square
of the ground state wavefunction at the position $r_{C}$ for which
the difference in the excited and ground state potentials equals the
photon frequency.

In our case the reflection approximation cannot be applied because
the asymptotic potentials in the ground and excited state are similar.
The ground-state wave function has nodes at 26.5 au, 32.4 au, and 46.1 au
for a scattering length $a_{scat}=389.8\;{\rm au}$ and a collision energy
$\varepsilon_{r}/k_{B}=1\;\mu{\rm K}$. For the reflection approximation
the relation between detuning and outer turning point of the excited
bound states is determined from a potential that is the sum of the
Hund's case (c) potential $V(0_u^+)$ and the rotational correction
$[2+J_{e}(J_{e}+1)]\hbar^{2}/(2\mu r^{2})$. The potential $V(0_u^+)$
approaches the $^3\Pi_u$ potential at large internuclear separation
and $J_e=1$. These ground state nodes correspond to outer turning
points detunings of $-$51.3 GHz, $-$15.0 GHz, and $-$1.6 GHz, respectively.
These detunings do not correspond with dark lines in Fig.~\ref{fig8}.

Appendix~\ref{appc} gives another perspective of Fig.~\ref{fig8}.
The data are described in terms of a near threshold vibrational
quantum number instead of the binding energy. Integer values of this
quantum number are related to the bins defined in Fig.~\ref{fig4}.
This discussion is not crucial for the main thrust of the paper and,
therefore, has been placed in an appendix.  It, however, gives a deeper
understanding of the physics involved and is worth presenting.

\begin{figure}
\includegraphics[angle=0,width=\columnwidth,clip]{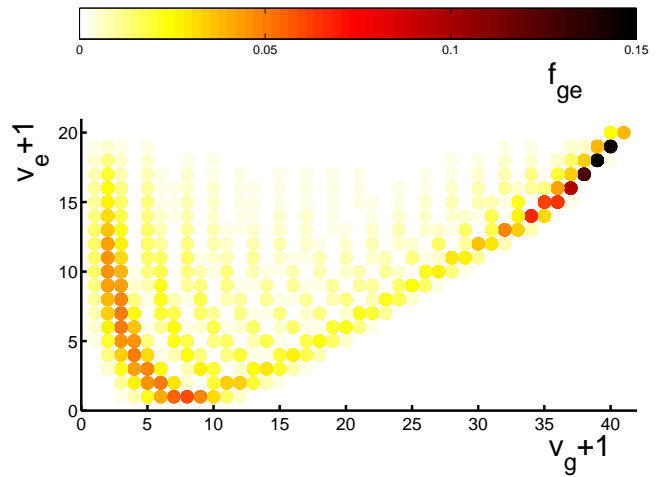}
\caption{The fraction $f_{ge}$ of
molecules in rovibrational state $v_g, J_g=0$ of the $^{1}\Sigma_{g}^+$
potential produced by natural decay of rovibrational state $v_{e},J_e=1$
of $0_u^+$ symmetry.  }
\label{fig9}
\end{figure}

Finally, we have investigated the possibility of creating cold molecules
in the ground electronic state via the photoassociation process. The
similar $C_6/r^6$ dependence of the ground and excited potentials
can make such a process more efficient than making ground state
molecules through excited states in which the asymptotic form of the
potential is $C_{3}/r^{3}$.  We have calculated rates for bound-bound
transitions between $0_u^+$ excited states and $^{1}\Sigma_{g}^+$
ground states. From these rates we have calculated the fraction of
ground-state molecules, which after photoassociation are formed by
natural decay.  Figure~\ref{fig9} presents the fraction $f_{ge}$ of
molecules in rovibrational state $v_g, J_g=0$ of the $^{1}\Sigma_{g}^+$
potential produced by natural decay of rovibrational state $v_{e},J_e=1$
of $0_u^+$ symmetry.
The fraction of a $v_g,J_g=0$ molecule is no more than 15\%.
For $v_e>10$ high $v_g$ vibrational levels of the ground state
potential are produced, while for $v_e<10$ low $v_g$ levels are produced.
Nevertheless, even for high $v_e$ a few percent of the ground-state molecules
have low $v_g$.
The fraction of molecules decaying to $J_{g}=2$ vibrational
states of the $^{1}\Sigma_{g}$ potential have also been calculated, but
are not shown. It was found that the fraction of excited molecules
decaying to the $J_{g}=2$ vibrational states is bigger than those
decaying to the $J_{g}=0$ vibrational states.
Rotational states with $J_{g}=1$ do not exist because of
Bose symmetrization (See appendix \ref{appa}).

Adding the $J_g=0$ and 2 fractions $f_{ge}$ of molecules shows that
most of the spontaneous emission of a $v_e,J_e$ $0_u^+$ rovibrational
level goes to ground molecular bound states.  In fact, for $v_{e}<19$
more than 80\% of the decay is to molecular states. The only exception
is the last vibrational level, $v_{e}=19$, corresponding to line 1 in
Fig.~\ref{fig5}(a), for which 30\% of the decay is to molecular states.
We have also found that the spontaneous decay rate of deeply-bound
excited states can be one order of magnitude larger than
$2\Gamma_{A}(^{3}P_{1})/\hbar$.

\section{Conclusions}

We have developed a description for the shape of photoassociation lines
for weak transitions in which the natural width is smaller than the
Doppler width.  The lineshape theory includes Doppler broadening and a
photon recoil shift.  It was shown that the Doppler effect significantly
affects the PA lineshape when the gas temperature is on the order of the
recoil temperature and below.

A model calculation for calcium has been carried out. It is an example of
photoassociation near the intercombination line of alkaline-earth atoms.
We find that photoassociation spectroscopy should be possible at calcium
gas temperatures on the order of $\mu$K and below.
In  addition, it was shown that when the long-range potential
of the ground and excited state are similar, the reflection approximation
\cite{bohn99,julienne96,boisseau00} incorrectly estimates the stimulated
width and strength of PA lines.
Finally, we have indicated that excited molecules are very likely to decay to
vibrational levels of the electronic ground state.  This can be used as
an effective way to produce cold molecules.

\acknowledgments

The authors wish to express their gratitude to A. Derevianko for making his
calculations of dispersion coefficients available before publication.
This work has been partially supported by the U.S. Office of Naval Research.
The research is part of the program of the National Laboratory FAMO in Toru\'n, Poland.

\appendix
\section{Close-coupling calculations}\label{appa}
\setcounter{equation}{0}

The quantum theory of slow-atom collisions \cite{mies73,gao96}
allows a quantitative description of the scattering process
and bound states. We apply this theory to describe the slow
collisions and  bound states of alkaline-earth atoms.

Scattering or bound states of two identical alkaline-earth
atoms without nuclear spin can be described in the basis
\begin{equation}
|SLjlJM_{J};p\rangle\equiv\sum_{m_{j},m_{l}}\langle
jl\,m_{j}m_{l}|JM_{J}\rangle |SLjm_{j};\sigma\rangle Y_{lm_l}(\theta,\phi)\,,
\label{a1}
\end{equation}
where the $|SLjm_{j};\sigma\rangle$ describe the electronic state of
the molecule and $Y_{lm}(\theta,\phi)$ is a spherical harmonic describing the
nuclear rotation. The quantity
$\vec{L}$ is the total electron orbital angular momentum,
$\vec{S}$ is the total electron spin angular momentum,
$\vec{\jmath}=\vec{L}+\vec{S}$ is the total electron angular momentum,
$\vec{l}$ is the rotational angular momentum,
and
$\vec{J}=\vec{l}+\vec{\jmath}$ is the total angular momentum.  The
projections of $\vec{\jmath}$, $\vec{l}$, and $\vec{J}$ on a space-fixed
$z$-axis are $m_j$, $m_l$, and $M_J$, respectively. The quantity $p$
is the total parity.  {\it Gerade} ($g$, $\sigma=+1$) and {\it ungerade}
($u$, $\sigma=-1$) electronic states correspond to total parity $p=+1$
and $p=-1$ states, respectively. This is a consequence of the more
general rule for atoms with nonzero nuclear spin $p=\sigma(-1)^{2I}$,
where $I$ is the atomic nuclear spin.  The total parity restricts
the allowed $l$ by $p=p_{A}p_{B}(-1)^{l}$, where $p_{A}$ and $p_{B}$
are the atomic parities.  The atomic parity is +1 for the ground $^1S$
state and $-$1 for excited $^1P$ and $^3P$ states.

The molecular Hamiltonian $H=T+ H_A+V_{\rm int}+V_{\rm rot}$ is calculated
in the $|SLjlJM_{J};p\rangle$ basis. Here, $T=-(\hbar^2/(2\mu)) d^2/dr^2$
is the kinetic energy operator, $H_A$ is the atomic Hamiltonian,
$V_{\rm int}$ are the nonrelativistic Born-Oppenheimer potentials,
and $V_{\rm rot}=\hbar^2 \vec{l}^2/(2\mu r^{2})$ describes the
rotational energy.  The matrix elements for the kinetic and rotational
energy are diagonal in this basis.  In fact, $\langle
SLjlJM_{J};p|T|SLjlJM_{J};p\rangle= -(\hbar^2/(2\mu)) d^2/dr^2$
and $\langle SLjlJM_{J};p|V_{\rm rot}|SLjlJM_{J};p\rangle= \hbar^2
l(l+1)/(2\mu r^{2})$.

The matrix elements for the Born-Oppenheimer potentials are calculated in
two steps. The first step involves transforming the molecular electronic
state $|SLjm_{j};\sigma\rangle$ into a body-fixed coordinate system. That
is into a superposition of $|SLj\Omega;\sigma\rangle$ states, where
$\Omega$ is the projection of $\vec{j}$ on the internuclear axis.
After some algebra the matrix elements are given by
\begin{eqnarray}
\lefteqn{\langle S'L'j'l'J'M_{J}';p'|V_{\rm int}|SLjlJM_{J},p\rangle=}
\nonumber\\
&&
\delta_{p',p}\delta_{J',J}\delta_{M_{J}',M_{J}}
\sqrt{\frac{(2l'+1)(2l+1)}{(2J+1)^2}}
\nonumber\\
&&
\sum_{\Omega}
\langle j'l'\Omega 0|J'\Omega\rangle
\langle jl\Omega 0|J\Omega\rangle
\langle S'L'j'\Omega;\sigma|V_{\rm int}|SLj\Omega;\sigma\rangle\,,
\nonumber\\
\label{a2}
\end{eqnarray}
i.e.  the operator $V_{\rm int}$ is diagonal in  $J$, $M_J$, and $p$ but
not diagonal in $j$ and $l$.  A similar transformation is discussed by
Napolitano {\it et al}. \cite{napolitano97} in the context of ultra-cold
collisions between atoms in the $^{1}S_{0}$ and $^{1}P_{1}$ states.

The next step is to express the body-fixed electronic states
$|SLj\Omega;\sigma\rangle =\sum_{\Sigma,\Lambda}\langle
SL\Sigma\Lambda|j\Omega\rangle |SL\Sigma\Lambda;\sigma\rangle$
in terms of $|SL\Sigma\Lambda;\sigma\rangle$, where $\Sigma$ and
$\Lambda$ are projections of $S$ and $L$ along the internuclear
axis. In other words,
\begin{eqnarray}
\lefteqn{\langle S'L'j'\Omega';\sigma'|V_{\rm int}|SLj\Omega;\sigma\rangle=}
\nonumber\\
&&
\sum_{\Sigma',\Lambda',\Sigma,\Lambda}
\langle S'L'\Sigma'\Lambda'|j'\Omega'\rangle
\langle SL\Sigma\Lambda|j\Omega\rangle \nonumber\\
&&
\quad\quad\quad
\langle S'L'\Sigma'\Lambda';\sigma'|V_{\rm int}|SL\Sigma\Lambda;\sigma\rangle \,.
\label{a3}
\end{eqnarray}
The Born-Oppenheimer potentials are diagonal in this
$|SL\Sigma\Lambda;\sigma\rangle$ basis
and the diagonal matrix elements are
$\langle SL\Sigma\Lambda;\sigma|V_{\rm int}|SL\Sigma\Lambda;\sigma\rangle=
V_{\rm int}(^{2S+1}|\Lambda|_{\sigma})$.

Finally, we calculate matrix elements of the Hamiltonian $H_{A}$ for two
non-interacting atoms, where one atom is always in the $^{1}$S$_{0}$ state
while the other atom can be in the state $^{1}$S$_{0}$, $^{1}$P$_{1}$,
or $^{3}$P$_{2,1,0}$.  Because one atom is in the $^{1}$S$_{0}$ state,
the molecular angular momenta $S$, $L$, and $j$ in Eq.~(\ref{a1})
are equivalent to those of the second atom.

A realistic description of the atomic Hamiltonian should include
relativistic coupling between singlet $^1P_1$ and triplet $^3P_1$
states.  Therefore, the atomic Hamiltonian is not diagonal in the
basis of Eq.~(\ref{a1}).  Following Mies {\it et al}.\cite{mies78} ``dressed''
electronic states for $j=1$, which are a mixture of singlet and triplet,
are introduced as follows
\begin{eqnarray}
\lefteqn{|\widetilde{S}=0,Ljm_{j};\sigma\rangle\equiv}
\nonumber\\
&&
\cos(\alpha)|S=0,Ljm_{j};\sigma\rangle+
\sin(\alpha)|S=1,Ljm_{j};\sigma\rangle
\nonumber
\end{eqnarray}
and
\begin{eqnarray}
\lefteqn{|\widetilde{S}=1,Ljm_{j};\sigma\rangle\equiv}
\nonumber\\
&&
-\sin(\alpha)|S=0,Ljm_{j};\sigma\rangle+
\cos(\alpha)|S=1,Ljm_{j};\sigma\rangle
\nonumber
\end{eqnarray}
where $\alpha$ is a small mixing angle.
For $j=0$ and 2 we have
$|\widetilde{S}Ljm_{j};\sigma\rangle=|SLjm_{j};\sigma\rangle$.
We assume that in this dressed basis the atomic Hamiltonian is diagonal
with diagonal matrix elements
$\langle\widetilde{S}LjlJM_{J};p|H_{A}|\widetilde{S}LjlJM_{J};p\rangle=
E_{A}(^{2\widetilde{S}+1}L_{j})$, where $E_{A}(^{2\widetilde{S}+1}L_{j})$
is the energy of the dressed state $^{2\widetilde{S}+1}L_{j}$ relative to
the $^1S_0$ ground state.

The mixing angle $\alpha$ is determined by the requirement that the
dressed basis $|\widetilde{S}Ljm_{j};\sigma\rangle$ reproduces the
experimental transition probabilities between the excited $^{1,3}$P$_{1}$
states and the ground $^{1}$S$_{0}$ state.  The angle can then be related
to the ratio of the experimental dipole moments of these transitions. In
fact, we used $\tan(\alpha)=\sqrt{[E_{A}(^{1}P_{1})/E_{A}(^{3}P_{1})]^{3}
[\tau_{A}(^{1}P_{1})/\tau_{A}(^{3}P_{1})]}$.  This approach is
only an approximation of the description of a real alkaline-earth
atom\cite{mies78}. For the purposes of this paper, however, it is
sufficient.  The values for the energies and lifetimes have been obtained
from Refs.~\cite{machholm01,wiese,drozdowski93} and listed in Table \ref{tab1}.
The experimental results on the $^{3}P_{1}$ lifetime are
compared by Drozdowski {\it et al}. \cite{drozdowski93} and
some uncertainty still exist. We have chosen the estimated
value of Ref. \cite{machholm01}, which lies between the experimental values.

\begin{table}
\caption{The values of energies and lifetimes of Ca atomic states used in this paper. }
\label{tab1}
\center
\begin{tabular}{c|rr}
\hline
\hline
$^{2S+1}L_{j}$ & $E_{A}(^{2S+1}L_{j})$ [cm$^{-1}$] & $\tau_{A}(^{2S+1}L_{j})$ \\
\hline
$^{1}P_{1}$ & 23 652.304 & 4.59 ns\\
$^{3}P_{2}$ & 15 315.943 & \\
$^{3}P_{1}$ & 15 210.063 & 0.48 ms \\
$^{3}P_{0}$ & 15 157.901 & \\
$^{1}S_{0}$ & 0.000 & \\
\hline
\hline
\end{tabular}
\end{table}

The molecular Hamiltonian $H$ conserves $J$, $M_{J}$, and $p$.
In fact, the matrix elements of $H$ in the basis of Eq.~(\ref{a1})
for given $J$ and $p$ are independent of $M_{J}$.  It is
convenient to introduce the channels $|\gamma\rangle=|SLjlJM_{J};p\rangle$,
$|\widetilde{\gamma}\rangle=|\widetilde{S}LjlJM_{J};p\rangle$, and note that
$\langle\widetilde{\gamma}'|H|\widetilde{\gamma}\rangle=
\sum_{\gamma',\gamma} \langle\widetilde{\gamma}'|\gamma'\rangle
\langle\gamma'|H|\gamma\rangle \langle\gamma|\widetilde{\gamma}\rangle$.
Close-coupling equations for the molecular wave function
$|\Psi\rangle=\sum_{\widetilde{\gamma}} |\widetilde{\gamma}\rangle
F_{\widetilde{\gamma}}(r)/r$ can be written as
\begin{equation}
\label{a4}
-\frac{\hbar^{2}}{2\mu}\frac{d^{2}}{dr^{2}}F_{\widetilde{\gamma}}(r)
+\sum_{\widetilde{\gamma}'}
\langle\widetilde{\gamma}|H_{A}+V_{\rm int}+V_{\rm rot}|\widetilde{\gamma}'\rangle F_{\widetilde{\gamma}'}(r)=
E\;F_{\widetilde{\gamma}}(r) \,,
\end{equation}
where $E$ is the total molecular energy.  These coupled Schr\"odinger
equations are solved numerically to find scattering and bound states.

The collision between two ground state atoms can be solved separately from
that of a ground plus an excited state atom.  Moreover, for photoassociation
of Ca we need scattering solutions of the Schr\"odinger equation for
ground state collisions and bound states for ground-excited molecules.

For the collision of two ground-state atoms $^{1}$S$_{0}$+$^{1}$S$_{0}$
there is only a single channel for given $J_g$, $M_{g}$ and $p_g$. The channel
is $|\gamma_{g}\rangle=|S_{g}L_{g}j_{g}l_{g}J_{g}M_{g};p_{g}\rangle$
with $S_{g}=0$, $L_{g}=0$, $j_{g}=0$ and $l_{g}=J_{g}$.  The two
$^{40}$Ca atoms are indistinguishable bosons and, therefore, the
wavefunction must be symmetric under exchange of atoms and only
channels with positive parity $p=+1$ exist.  Consequently, only even
$l_g=0,2,4,\dots$ are allowed and the atoms interact on the gerade
electronic state $^{1}\Sigma^{+}_{g}$.  Solving the Schr\"{o}dinger
equation with $E=\varepsilon_{r}>0$ we obtain the scattering
wavefunction $|\Psi_{g}^{+}(\varepsilon_{r},J_{g},M_{g})\rangle=
|\gamma_{g}\rangle F_{\gamma_{g}}^{+}(r)/r$.
This wavefunction occurs in the stimulated width defined
by Eq.~(\ref{e5}).  For large $r$ the wavefunction goes to
$F_{\gamma_{g}}^{+}(r)\rightarrow \sqrt{2\mu/(\pi\hbar^{2} k_{r})}
\sin(k_{r}r+\pi l_{g}/2+\eta_{l_{g}})\exp(i\eta_{l_{g}})$ \cite{julienne96,weiner99}.
Our interest will be in the solution for $J_{g}=l_{g}=0$ or $s$-wave
collisions.  For collisions at ultralow temperatures other partial
waves do not contribute significantly.

For molecules formed by a ground- and excited-state atom there
are multiple channels involved. The number of channels is
determined by $J_e$ and $p_e$. Photon selection rules
limit the allowed total angular momentum of the excited bound states
to $J_e=J_g,J_g\pm1$ and their parity to $p_e=-p_g$. For $J_g=0$ or
$s$-wave collisions PA can only make $J_e=1$ and $p_e=-1$ bound states.
Table~\ref{tab2} lists the five channels for $J_e=1$ and $p_e=-1$.
The first two channels correspond to $^{1}$S$_{0}$+$^{3}$P$_{1}$ states.
There are no channels with $j_{e}=0$ and, since the atomic
energies satisfy  $E_A(^3P_0)<E_A(^3P_1)<E_A(^3P_2)$, predissociation
of bound states below the $^1S_0+^{3}P_{1}$ limit does not occur.
In fact, this is true for all odd $J_{e}$ and $p_{e}=-1$.

\begin{table}
\caption{The allowed $J_e=1$, $p_e=-1$ and $M_{e}=0,\pm1$ dressed channels
of interacting ground- and excited-state alkaline-earth atoms. }
\label{tab2}
\center
\begin{tabular}{c|cccccc}
\hline
\hline
$|\widetilde{\gamma}_{e}\rangle$ & $\widetilde{S}_{e}$ & $L_{e}$ & $j_{e}$ & $l_{e}$ \\
\hline
1 & 1 & 1 & 1 & 0 \\
2 & 1 & 1 & 1 & 2 \\
3 & 1 & 1 & 2 & 2 \\
4 & 0 & 1 & 1 & 0 \\
5 & 0 & 1 & 1 & 2 \\
\hline
\hline
\end{tabular}
\end{table}

The numerical solutions of the Schr\"odinger equation for the
scattering collision in the ground state are calculated
using the Numerov method \cite{johnson77,johnson78} implemented
in the close coupling code developed by Mies, Julienne and Sando
\cite{miesCC}. The coupled-channel bound states calculations
for the excited state are carried out using the discrete variable
representation (DVR) \cite{dulieu95,colbert92} following
Tiesinga {\it et al}. \cite{tiesinga98}.

\section{Interaction with light}\label{appb}
\setcounter{equation}{0}

In our treatment of the photoassociation process we need the
stimulated width, defined by Eq.~(\ref{e4}),
between the scattering ground state and the excited bound states.
The stimulated width can be expressed as
\begin{equation}
\label{b1}
\Gamma_{eg}(\varepsilon_{r})=2\pi
\left|
\sum_{\widetilde{\gamma}_{e}}
\int_{0}^{\infty}dr\;
\langle\widetilde{\gamma}_{e}|V_{\rm las}|\gamma_{g}\rangle
F_{\widetilde{\gamma}_{e}}^{*}(r)F_{\gamma_{g}}^{+}(r)
\right|^{2}\,,
\end{equation}
where the wavefunctions are expressed in terms of channel functions
$F_{\widetilde{\gamma}_{e}}(r)$ and $F_{\gamma_{e}}^{+}(r)$ and
$\langle\widetilde{\gamma}_{e}|V_{\rm las}|\gamma_{g}\rangle$ are matrix
elements between channels in the ground and excited state.  If we only
consider dipole transitions and neglect retardation effects, this matrix
element is independent of the interatomic separation $r$ and given by
\begin{equation}
\label{b2}
\langle\widetilde{\gamma}_{e}|V_{\rm las}|\gamma_{g}\rangle=
\sqrt{\frac{2\pi I}{c}}\frac{1}{\sqrt{4\pi \epsilon_{0}}}
\langle\widetilde{\gamma}_{e}|\vec{d}\cdot \vec{e}_q |\gamma_{g}\rangle
\end{equation}
where the laser has intensity $I$ and polarization
$\vec{e}_q$, $\epsilon_{0}$ the permittivity of vacuum, and
$\langle\widetilde{\gamma}_{e}|\vec{d}\cdot \vec{e}_q |\gamma_{g}\rangle=
\sum_{\gamma_e} \langle \widetilde{\gamma}_{e}|\gamma_e\rangle
\langle\gamma_{e}|\vec{d}\cdot \vec{e}_q |\gamma_{g}\rangle $
is the molecular dipole matrix element between spin channels.

Napolitano {\it et al} \cite{napolitano97} discuss the connection between
molecular and atomic dipole matrix elements. A similar approach is used
here to find matrix elements.
The molecular dipole matrix elements between ground and excited state are
\begin{eqnarray}
\lefteqn{\langle S_{e}L_{e}j_{e}l_{e}J_{e}M_{e};p_{e}|d^{K}_{q}|S_{g}L_{g}j_{g}l_{g}J_{g}M_{g};p_{g}\rangle=}
\nonumber\\
&&
\sqrt{\frac{(2l_{e}+1)(2l_{g}+1)}{(2J_{e}+1)^2}}
\langle J_{g}KM_{g}q|J_{e}M_{e}\rangle
\nonumber\\
&&
\sum_{\Omega_{g},\Omega_{e},q'}
\langle J_{g}K\Omega_{g}q'|J_{e}\Omega_{e}\rangle
\langle j_{e}l_{e}\Omega_{e} 0|J_{e}\Omega_{e}\rangle
\langle j_{g}l_{g}\Omega_{g} 0|J_{g}\Omega_{g}\rangle
\nonumber\\
&&
\langle S_{e}L_{e}j_{e}\Omega_{e};\sigma_{e}|d^{K}_{q'}|S_{g}L_{g}j_{g}\Omega_{g};\sigma_{g}\rangle
\label{b3}
\end{eqnarray}
where the spherical tensor operator $d^K_q=\vec{d}\cdot \vec{e}_q$
with $K=1$ and $q=-1,0$, or $+1$. In Eq.~(\ref{b3}) we have
expressed the dipole operator in a body-fixed coordinate system.
For both singlet $S=0$ ground and excited states we realize that
$|SLj\Omega;\sigma\rangle=|SL\Sigma\Lambda;\sigma\rangle$
and thus $j=L$ and $\Omega=\Lambda$.  It then follows
\begin{eqnarray}
\lefteqn{
\langle S_{e}L_{e}j_{e}\Omega_{e};\sigma_{e}|d^{K}_{q}|S_{g}L_{g}j_{g}\Omega_{g};\sigma_{g}\rangle =}
\nonumber\\
&&
\langle S_{e}L_{e}\Sigma_{e}\Lambda_{e};\sigma_{e}| d^{K}_{q}|
S_{g}L_{g}\Sigma_{g}\Lambda_{g};\sigma_{g}\rangle\,,
\label{b4}
\end{eqnarray}
for electronic states with {\it zero} total electron spin $S_e=S_g=0$.
In the $|SL\Sigma\Lambda;\sigma\rangle$ basis the dipole operator can
be evaluated in terms of the atomic linewidth, $\Gamma_{A}(^{1}P_{1})$,
of the $^1P_1$ atomic state decaying to the ground $^1S_0$ state.
There are four distinct matrix elements
$\langle 010\Lambda;\sigma|d^{1}_{\Lambda}| 0000;g\rangle=d(^{1}|\Lambda|_{\sigma})$
given by
\begin{eqnarray}
d(^{1}|\Lambda|_{\sigma})&=&
\sqrt{
\frac{3}{4}
\left(\frac{\lambda_{A}(^{1}P_{1})}{2\pi}\right)^{3}
4\pi \epsilon_{0}\Gamma_{M}(^{1}|\Lambda|_{\sigma})
}\,,
\label{b5}
\end{eqnarray}
where $\Gamma_{M}(^{1}|\Lambda|_{\sigma})$ is the molecular linewidth.
The $d(^1|\Lambda|_\sigma)$ are $r$-independent quantities as
we neglect retardation \cite{machholm01,meath68,power67} and,
therefore, the molecular linewidth can be well approximated by
$\Gamma_{M}(^{1}|\Lambda|_{u})=2\Gamma_{A}(^{1}P_{1})$ for ungerade
states and $\Gamma_{M}(^{1}|\Lambda|_{g})=0$ for gerade states.  In our
phase convention the $d(^1|\Lambda|_\sigma)$ are positive. The molecular
dipole matrix elements between the triplet $S_e=1$ and the ground $S_g=0$
states are zero.

\section{Near threshold vibrational quantum number}\label{appc}
\setcounter{equation}{0}

Following LeRoy and Bernstein \cite{leroy70} the JWKB quantum condition
for the eigenvalues of energies $E$ for a potential $V(r)$ is that
$\Phi(E)=v\pi$ where $v$ is an integer and the phase
$\Phi(E)=\Phi_{JWKB}(E)-\pi/2$. Here
$\Phi_{JWKB}(E)=\int_{r_{1}}^{r_{2}}dr\;\kappa(E,r)$
is the semiclassical phase integral of
$\kappa(E,r)=\sqrt{(E-V(r))2\mu/\hbar}$ calculated between
the classical turning points $r_{1}$ and $r_{2}$.
This criterion is valid for deeply
bound states but breaks down near the dissociation
limit\cite{boisseau98}. It can be shown that for energies $E$
close to the threshold this condition still holds \cite{gribakin93}
if the phase $\Phi(E)$ is modified in the following way:
$\Phi(E)=\Phi_{JWKB}(E)-\pi/2-\pi/(2q-4)$ for long-range
potentials of the form $V(r)=C_{q}/r^{q}$.
A recent discussion on improvements to the LeRoy-Bernstein approach
can be found in Ref.~\cite{comparat04}.

The phase $\Phi(E)$ allows us to define a {\it generalized vibrational
quantum number} as $(\Phi(E)-v_{0}\pi)/\pi$, a continuous function of
energy $E$, where $v_{0}$ is an arbitrary constant, which does not need
to be an integer.  For $v_{0}=0$ the generalized vibrational quantum
number is 0, 1, ... ,$v_{\rm max}-1$, or $v_{\rm max}$ when $E$ is
equal to the energy of a bound state. Here, a zero value corresponds
to the most deeply bound state and $v_{\rm max}$ is the vibrational
quantum number of the last bound state. At the dissociation limit this
generalized vibrational quantum number equals $v_{D}=\Phi(0)/\pi$ and, in
general, can have a non-integer value\cite{leroy70}.

In the analysis of bound states near the threshold it is convenient to set
$v_{0}=v_{D}$ or $v_{0}=v_{\rm max}+1$. In the first case the generalized
vibrational quantum number is equal to zero at threshold. Negative
integers -1, -2, -3, ... correspond to energies that mark the edges of
``bins'', in which there is exactly one bound state.  Only for integer
$v_{D}$, however, ``bin edges'' coincide with bound state energies of
the potential. In fact, ``bin edges'' were introduced in this way in
Fig.~\ref{fig4}.
A second useful choice of $v_0$ is $v_{0}=v_{\rm
max}+1$. It allows us to define a {\it near threshold vibrational quantum
number}, which has values -1, -2, -3, ... for bound states counting from
the top.  The near threshold vibrational quantum number for
$E=0$ is $v_{D}-v_{max}-1$, and lies between -1 and 0.

\begin{figure}
\includegraphics[angle=0,width=\columnwidth,clip]{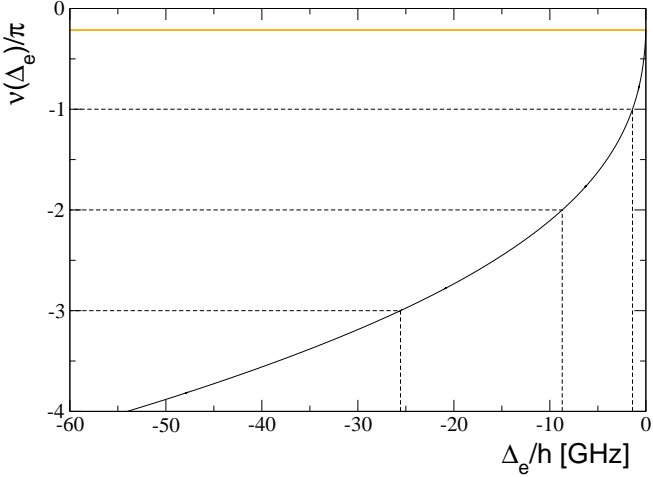}
\caption{The near threshold vibrational quantum number
$\nu(\Delta_{e})/\pi$ of the potential $V(^{3}\Pi_{u})+
[2+J_{e}(J_{e}+1)]\hbar^{2}/(2\mu r^{2})$ with $J_e=1$ as a function
of the binding energy $\Delta_{e}$.  The dashed lines indicate the
binding energies at which $\nu(\Delta_{e})/\pi$ is integer.  The value
of $\nu(\Delta_{e})/\pi$ at $\Delta_{e}=0$ is also indicated.}
\label{fig10}
\end{figure}

In practice, for energies far from threshold but not far enough to justify
the use of the JWKB approximation the energy dependence of the effective
vibrational quantum number can be calculated using expressions given
by Mies\cite{mies84}. Following Refs.~\cite{mies84,mies84j} we define a
near threshold vibrational quantum number $\nu(E)/\pi$ as a continuous
function of energy $E$, where $\nu(E)$ is the phase difference between
two solutions of the Schr\"odinger equation $f(r)$ and $\phi(r)$ at the
equilibrium separation, $r_{e}$, of the potential $V(r)$. The function
$f(r)$ is obtained by solving the Schr\"odinger equation
assuming that $f(0)=0$.  The function $\phi(r)$ is obtained by
solving the Schr\"odinger equation
assuming that $\phi(\infty)=0$. Then $\tan(\nu(E))$ can be calculated
from \cite{mies84}
\begin{equation}
\tan(\nu(E))=\frac{\kappa(E,r_{e})[f(r_e)\phi'(r_{e})-\phi(r_e)f'(r_{e})]}
{\kappa^{2}(E,r_{e})f(r_e)\phi(r_{e})+\phi'(r_e)f'(r_{e})}\,,
\label{c1}
\end{equation}
where the primes denote the first derivative with respect to $r$.
The near threshold vibrational quantum number, defined this way, has
an integer value for energies corresponding to the bound states of
the potential $V(r)$.

\begin{figure}
\includegraphics[angle=0,width=\columnwidth,clip]{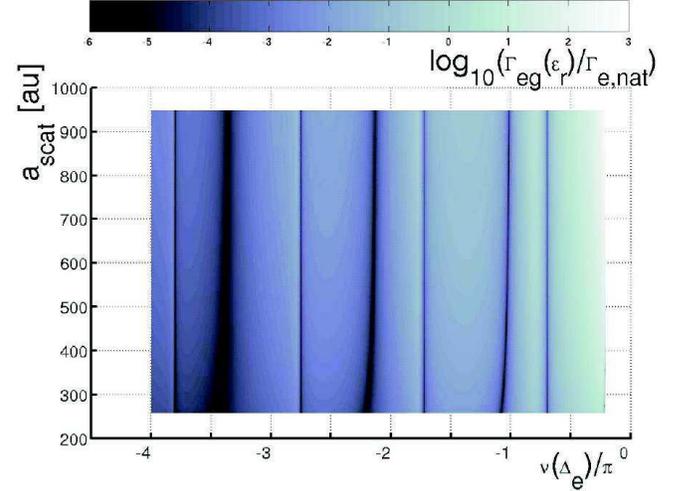}
\caption{The stimulated linewidth of the $0_u^+$ band as a function of
the near threshold vibrational quantum number $\nu(\Delta_{e})/\pi$
and the scattering length $a_{\rm scat}$ of the ground state potential.
The laser intensity is 1 W/cm$^2$ and the kinetic energy
is $\varepsilon_{r}/k_{B}=1\;\mu{\rm K}$.}
\label{fig11}
\end{figure}

A connection between the near threshold vibrational quantum number
and the binding energies $\Delta_e$ of the multichannel excited bound
states, used in Fig.~\ref{fig8}, can be made using quantum
defect theory\cite{mies84}. The theory states that the {\it shape} of the
energy dependence of $\nu(E)$ near threshold is nearly independent of the
short-range form of the potential and that we can replace the energy $E$
in Eq.~\ref{c1} by the binding energy $\Delta_e$.

The near threshold vibrational quantum number $\nu(\Delta_e)/\pi$ for a
potential $V(^{3}\Pi_{u})+ [2+J_{e}(J_{e}+1)]\hbar^{2}/(2\mu r^{2})$ as a
function of $\Delta_e$ is shown in Fig.~\ref{fig10}.  For this potential
the short range shape is adjusted to obtain {\it single-channel} bound
states (or negative integer near threshold vibrational quantum numbers)
for binding energies equal to -1.39 GHz, -8.71 GHz, -25.56 GHz and -54.01
GHz.  These values agree well with the {\it multi-channel} ``bin'' edges
-1.45 GHz, -8.75 GHz, -25.56 GHz and -53.94 GHz shown in Fig.~\ref{fig4}.
In fact, for binding energies below -0.026 GHz the multi-channel $0_u^+$
bound states are well approximated by this single-channel Hund's case(c)
potential. The connection between integer $\nu(\Delta_{e})/\pi$ and
the multi-channel bin edges breaks down for smaller binding energies as
Coriolis mixing changes the multi-channel coupling scheme from Hund's
case (c) to (e).  Since a single-channel can not fully simulate the
multi-channel bound states, the near threshold vibrational quantum number
$\nu(\Delta_{e})/\pi=-0.2135$ is not integer for $\Delta_e=0$.

Figure~\ref{fig11} shows the results of Fig.~\ref{fig8} in terms of the
near threshold vibrational quantum number $\nu(\Delta_{e})/\pi$ instead
of the binding energy $\Delta_e$.  It is clearly seen from Fig.~\ref{fig11}
that for $\nu(\Delta_{e})/\pi$ between $v$ and $v+1$, where $v=-4,-3,\;{\rm or}\;-2$,
and thus for each corresponding bin a pair of dark structures occurs.
In fact, the ``first kind'' of structure always appears at nearly the
same location within a bin.  The ``second kind'' of structure moves much
more within a bin.  Both observations reflect the fact that the $C_{6}$
coefficients of the ground and exited state potentials are similar.

\end{document}